\documentclass[final,1p,times]{elsarticle} 
\usepackage{graphicx} 
\usepackage{amssymb} 
\usepackage{amsthm} 
\usepackage{lineno} 

\journal{Nuclear Physics A} 
\begin{document} 

\begin{frontmatter} 


\title{Photoproduction at the Relativistic Heavy Ion Collider with STAR}

\author{Yury Gorbunov for the STAR collaboration}

\address{Physics Department, Creighton University, 2500 California plz, Omaha, NE, 68178}

\begin{abstract} 

The strong electromagnetic fields of short duration associated with
relativistic heavy ions make a heavy-ion collider a unique tool to study two-photon and photonuclear collisions.
 In this talk, we introduce the principles of photoproduction at hadron
  colliders, review  recent results from RHIC on meson and e$^+$e$^-$ production. 
 At RHIC, STAR has studied  exclusive $\rho^{0}$ vector meson
  production and $\rho^{0}$ production  accompanied by electromagnetic dissociation of both nuclei  in  collisions
  of AuAu at 62, 130 and 200 GeV. Recent results suggest the validity of the Glauber calculations for the vector
meson photoproduction and incosistency of the model based on the parton saturation phenomenon. The measurements are  also sensitive to 
 interference  between production on the two nuclei: either ion can be the photon
  emitter or the target. The level of observed interference suggests that the final state wave function carries
information about all possible decays long after the decay occurs.  We also observe coherent photoproduction of a $\pi^{+}\pi^{-}\pi^{+}\pi^{-}$
  state which may be associated with $\rho^{0*}$ (1450).

\end{abstract} 

\end{frontmatter} 






\section{Introduction}\label{intro}

Photoproduction can be observed in  ultra-peripheral heavy ion collisions  (UPCs), which  occur when  the impact
parameter  $b$ is  more than  twice the  nuclear radius  $R_A$ and an electromagnetic field of one nucleus
interacts   with  another  nucleus~\cite{baur}.  The electromagnetic  field of a relativistic nucleus may  be
represented as  a flux of  almost-real virtual photons,  following the
Weizs\"{a}cker-Williams method~\cite{ww}.  The photon flux scales as
the  square of the  nuclear charge  and so  the cross-sections  can be
large in heavy ion interactions. These photons can fluctuate into a
quark-antiquark pair, which can scatter elastically from the other
nucleus, emerging as a real vector meson. 

Also, purely electromagnetic interactions can take place. 
These processes can be described as two-photon interactions (in most cases) at 
energy scales below\ \ $\hbar/R_A$. 

Processes with multiple photon exchanges are possible which may excite the target nuclei into a giant
dipole resonance or higher excitation states. When excited nuclei decay they emit one or more neutrons.

\section{Coherent and Incoherent Photoproduction}\label{crosssection}

\begin{figure}[htb]
\centering
\includegraphics[scale=0.25]{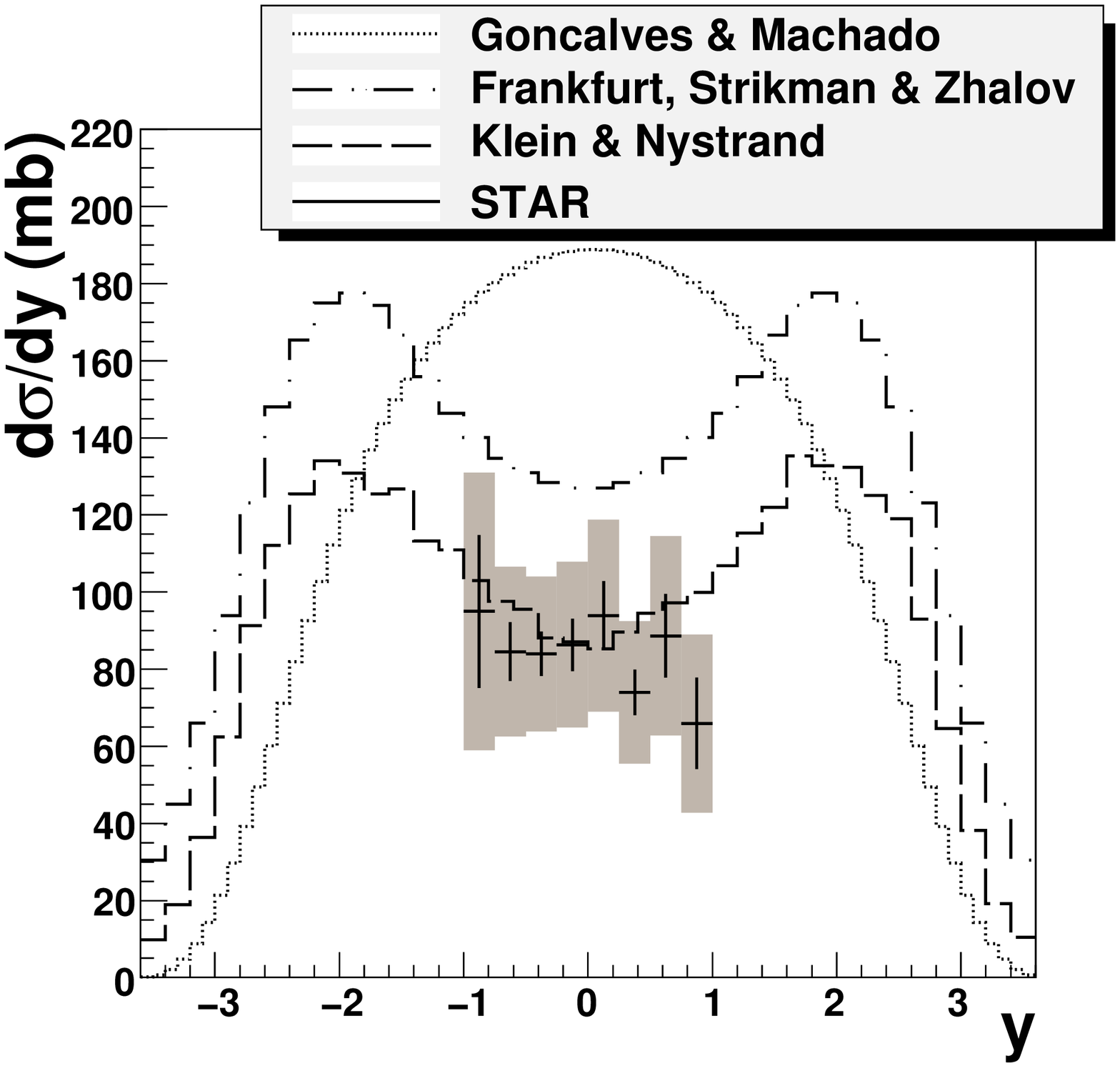}
\includegraphics[scale=0.42]{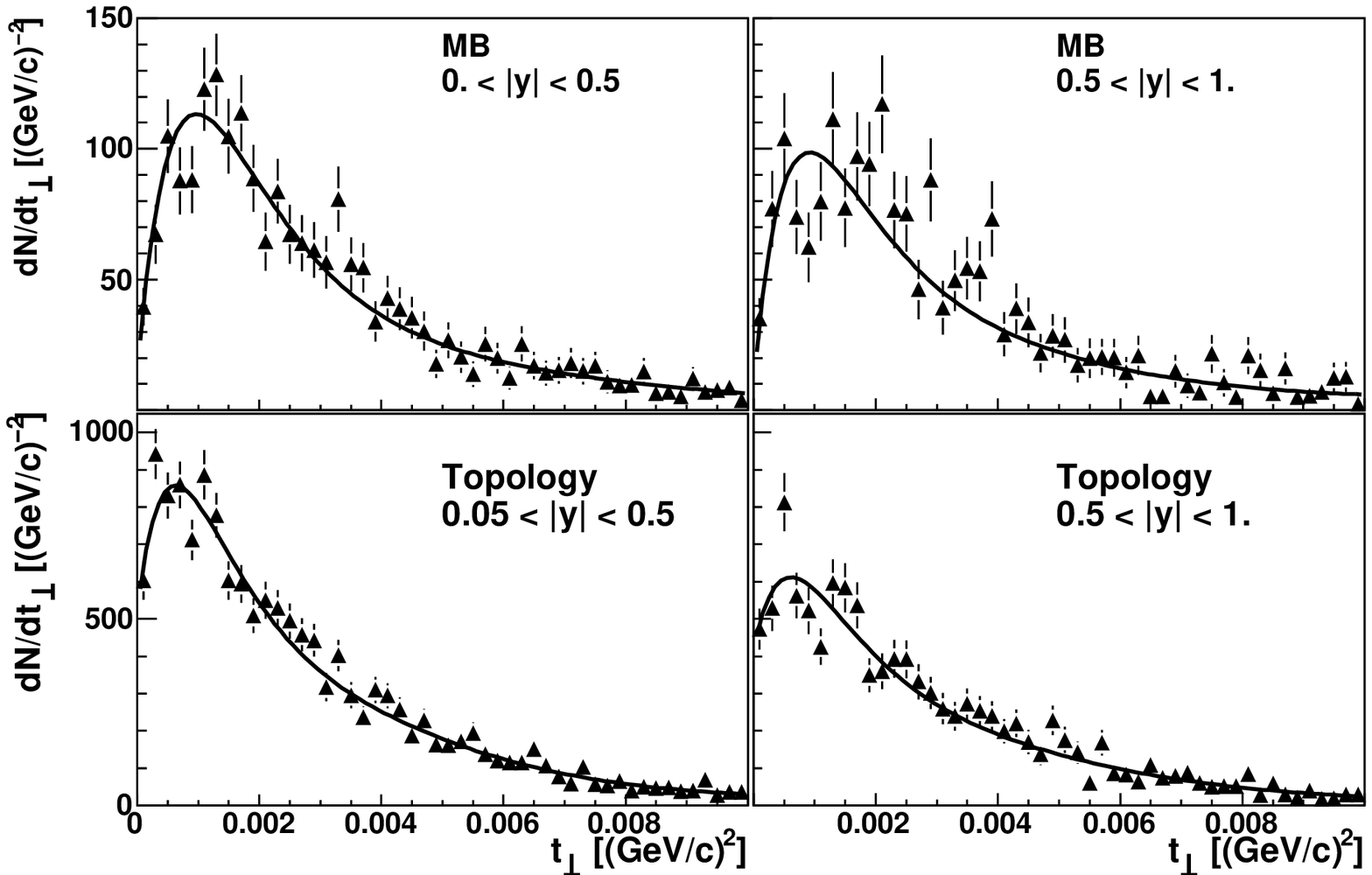}
\caption{\label{finalcross} Left: Comparison of theoretical predictions (dotted - QCD  
color dipole approach~\cite{gm}, dash dotted - generalized quantum VMD and the QCD Gribov-Glauber approach~\cite{fsz}, dashed - vector meson dominance plus a
classical mechanical Glauber approach~\cite{ksjn}) to the measured differential
cross-section for coherent $\rho^{0}$
production. The statistical errors are shown by the solid vertical line at each data point.
The sum of the statistical and systematic error bars is shown by the grey band. 
Right: Efficiency corrected $t_\perp (\approx p_T^2)$ spectrum for $\rho^0$ from
 minimum bias (top) and  topology data (bottom), for mid-rapidity (left)
and larger rapidity (right) samples.   The points are the
data, while the solid lines are the results of fits to Eq.~\ref{eq:fit}.
}
\end{figure}

Several data sets were used to study the production cross section of  
$\rho^0$ mesons at $\sqrt{s_{NN}}$ = 62, 130 and 200~GeV~\cite{STARrho^0}.  The STAR detector description, 
trigger setup and data analysis details can be found elsewhere~\cite{STARrho^0,tpcdes,zdcdes,trigdes}.
Recent results obtained with data collected at $\sqrt{s_{NN}}$ = 200~GeV allowed a detailed 
comparison between measured cross section and theoretical 
prediction~\cite{ksjn,fsz,gm}. The cross section comparison is shown in left panel of Fig.~\ref{finalcross}.





Due  to the  narrow  acceptance in  rapidity,  we cannot  distinguish
between those theoretical  models based on the shape. However, the measured 
cross section value can be used  to eliminate  models  which significantly
overestimate  the  total  production  cross-section  in  the  measured
rapidity range.

During the year 2004 run STAR took data of AuAu collisions at $\sqrt{s_{NN}}$ = 62~GeV. 
The data set was accumulated with minimum bias trigger and contains mainly coherently 
produced $\rho^0$ accompanied by the mutual excitation. The event selection and analysis 
procedure closely follow the one outlined in~\cite{STARrho^0}.
 The measured  $\rho^0$ coherent production cross section with mutual excitation was compared with theoretical 
prediction and found to be in agreement with the model described in~\cite{ksjn}.

\subsection{Interference}\label{interferencexx}

As was mentioned earlier the photoproduction can occur at large impact parameter $b$. For 
the $\rho^0$ photoproduction the median impact parameter $\langle b\rangle$ is about 46 fm (topology triggered data) which means that 
the $\rho^0$ source consists of two well-separated nuclei. As the result there are two indistinguishable possibilities:
either nucleus 1 emits a photon which scatters off nucleus 2, or
vice versa so that the system acts like a 2-slit interferometer with slit separation $b$. Since the produced
vector mesons have negative parity, the two amplitudes combine with opposite signs and as the result the cross section is  \cite{interfere}
\begin{equation}
\sigma(p_T,b,y)\! =\! \bigg| A(p_T,b,y) - A(p_T,b,-y)\exp{(i\vec{p}_T\cdot \vec{b})}\bigg|^2,
\label{eq:sigmay}
\end{equation}
where $A(p_T,b,y)$ and $A(p_T, b,-y)$ are the amplitudes for $\rho^0$ production at rapidity $y$ and transverse 
momentum $p_T$  from the two photon directions. 

The produced $\rho^0$s decay almost immediately at two well-separated points, so
any interference must develop after the decay, and involve the $\pi^+\pi^-$ final state.
Since the pions go in different directions, this requires an entangled $\pi^+\pi^-$ wave function
which cannot be factorized into separate $\pi^+$ and $\pi^-$ wave functions.  A measurement of
the two-source interference is sensitive to any loss of quantum mechanical coherence,
be it due to interactions with the environment \cite{decoherencereview} or as
a characteristic of the $\rho^0$ decay.

%

Figure~\ref{finalcross} right shows the efficiency corrected  minimum bias (MB) and topology data.
All four distributions show a dip at small $t_\perp (\approx p_T^2)$. The dip at low $t_\perp$ is broader 
for the MB data because $\langle b\rangle$ is smaller. The suppression at 
$t_\perp=0$  is larger for the small-rapidity samples
because the amplitudes for the two photon directions are more similar.

The $dN/dt_\perp$ spectrum is fit by the 3-parameter form
\begin{equation}
{dN\over dt_\perp} = A \exp(-kt_\perp) [1+ c(R(t_\perp)-1)],
\label{eq:fit}
\end{equation}
where $R(t_\perp)$  is the ratio of the simulated $t_\perp$-spectra with and
without interference, $A$ is an overall (arbitrary) normalization and 
$c$ gives the degree of spectral modification. 

We measured the interference to be at  $87\pm5$(stat.)$\pm8$ (syst.)\% of the expected level~\cite{inter_rho}.
This demonstrates that the final state wave function retains amplitudes for all 
possible decays, long after the decay occurs.

\subsection{$\pi^+\pi^-\pi^+\pi^-$ Production}\label{4prongx}

During the 2004 and 2007 runs STAR took data in AuAu colisions at  
$\sqrt{s_{NN}} = 200$~GeV with modified version of the minimum bias 
trigger. The trigger has been modified to accept events with 
multiplicity similar to events with four charged primary tracks. 

 The measurement of diffractive photoproduction of  
charged four-pion states available right now are mostly from 
$\gamma p$ and $\gamma d$ fixed target experiments at photon energies below 10 GeV. 
The heaviest nucleus used up to this point was carbon and photon energies ranging from 50 
to 200 GeV~\cite{ref4prong}. The experiments observed a broad peak at the order of 200 MeV 
in the range from 1430 to 1500 MeV/c$^2$.  One possible explanation is that this distribution is caused 
by the overlap of two separate states $\rho(1450)$ and $\rho(1700)$.

In the data analysis events with four charged tracks with net charge zero 
 and with total transverse momentum below 150~MeV/c have been selected. The cut on the 
transverse momentum ensures that selected events have been coherently photoproduced.

\begin{figure}[htb]
\centering
\includegraphics[scale=0.30]{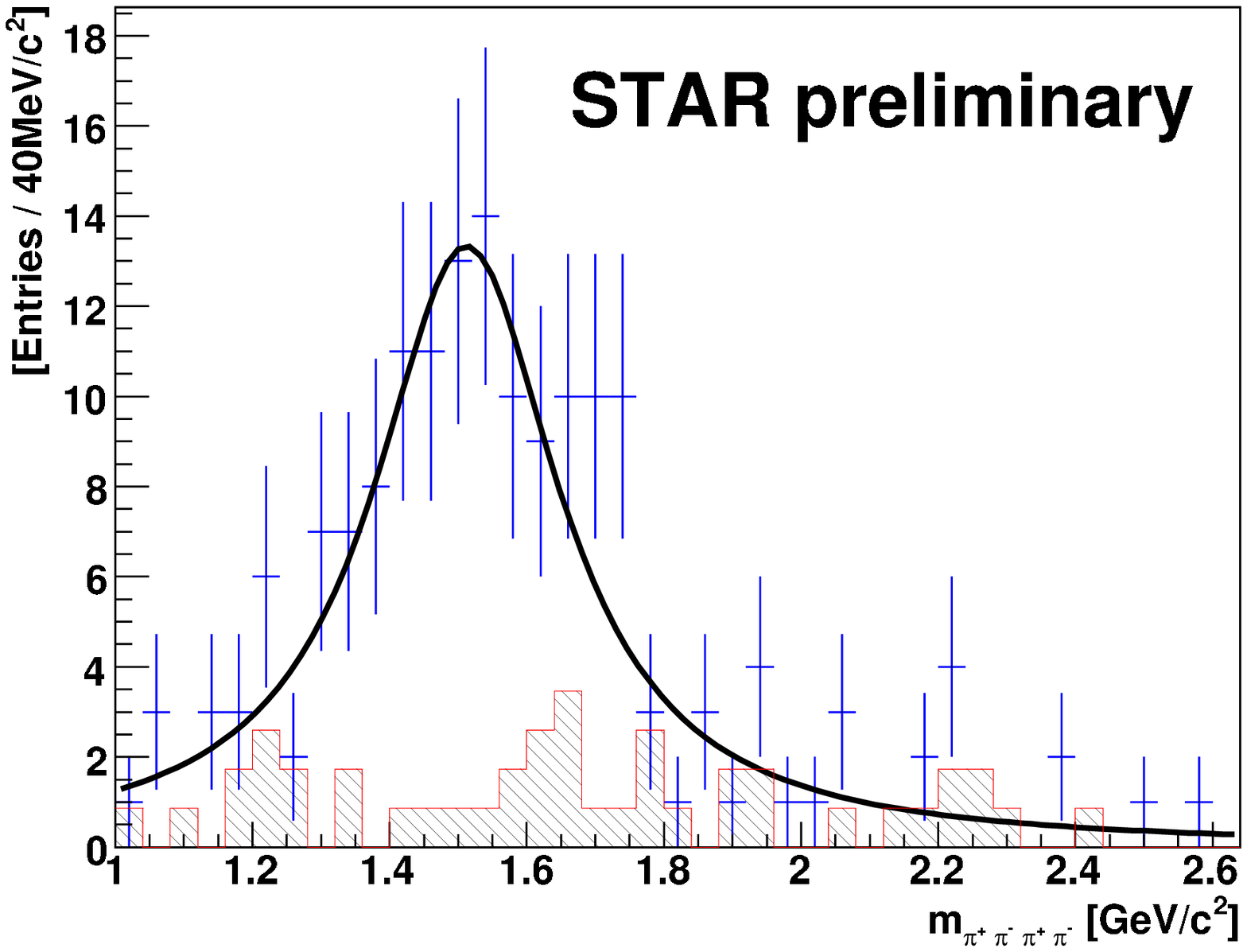}
\includegraphics[scale=0.30]{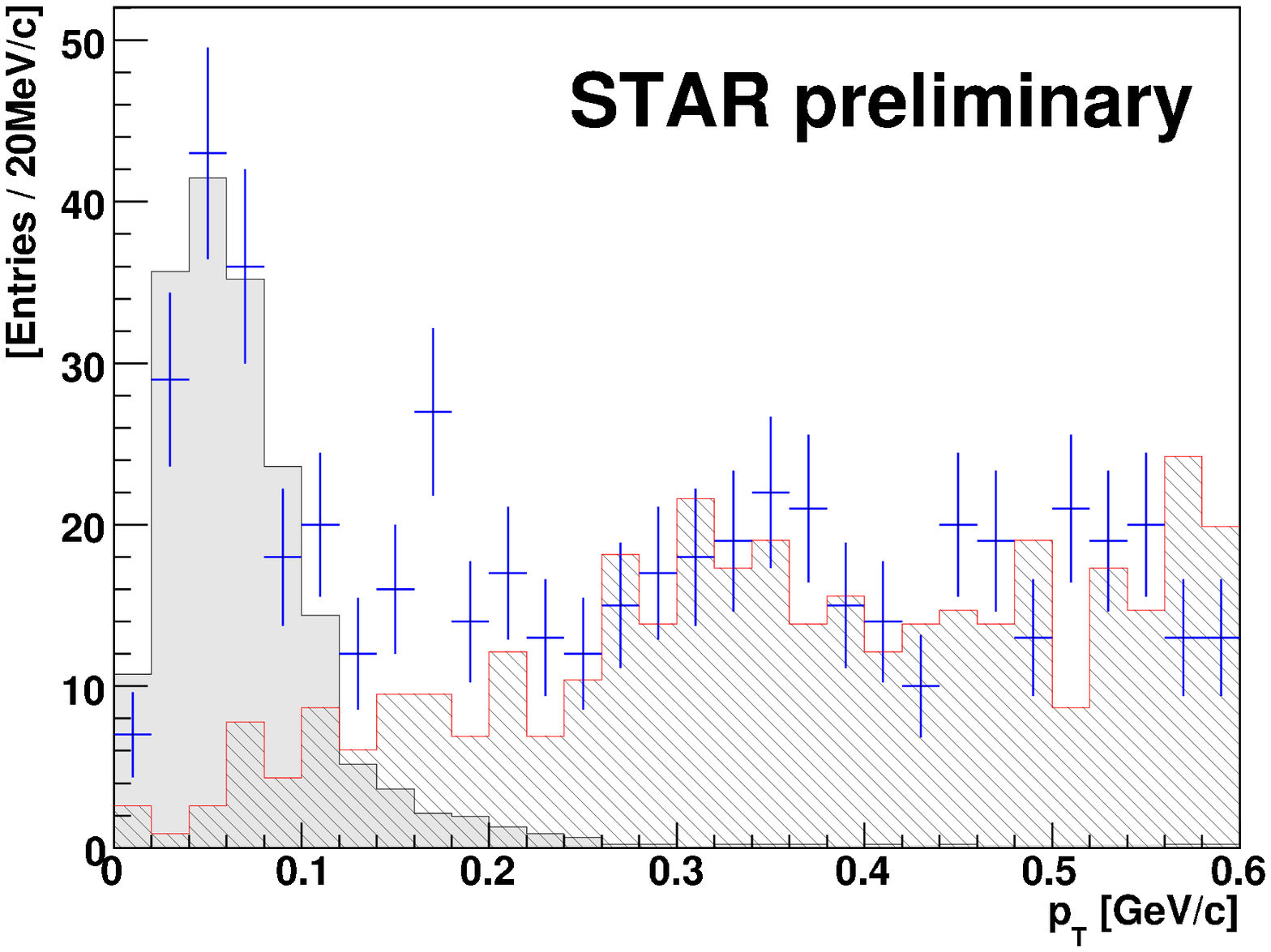}

\caption[]{Left plot shows invariant mass distribution for coherently produced $\pi^+\pi^-\pi^+\pi^-$ state. 
Right plot shows  $p_T$ distribution of coherently produced $\pi^+\pi^-\pi^+\pi^-$. The hatched distribution histogram  
shows background distribution estimated with four charged tracks with total non zero net charge. Shaded histogram 
shows signal distribution obtained with simulation.}
\label{4prong}
\end{figure}

As can be seen in Fig.~\ref{4prong} the transverse momentum distribution peaks at low $p_T$, typical for the coherent production. 
The Breit-Wigner gives a resonance mass of 1510 $\pm$ 20 MeV/c$^2$ and a width of 330 $\pm$ 45 MeV. The invariant mass distribution consist 
of approximately 100 candidates~\cite{boris}.

Previous fixed target experiments observed $\rho^\prime$ as an enchancement in 
the $\rho^\prime \rightarrow \pi^+\pi^-$ decay mode. STAR data show no significant  enhancement 
around 1500 MeV/c$^2$ in the double pion decay mode.

\section{Conclusions}\label{concl}

Photoproduction  of $\rho^{0}$ mesons  has been  measured in  the STAR
detector   at   RHIC  in Au-Au collisions   at
$\sqrt{s_{NN}}$   =  62, 130, 200~GeV.   Coherent  and   incoherent  $\rho^{0}$
photoproduction  has  been  observed. The production of $\rho^{0}$ mesons  is observed  
with and without accompanying Coulomb nuclear   excitations.  

The differential cross section has been measured as the function of rapidity,  
invariant mass and $t_\perp$ which allowed to isolate the incoherent part of the $\rho^0$ production. The 
measured total production cross section has been compared to the available theorectical models.

The interference effect has been observed in the $\rho^0$ production and found to be at the 
level of  $87\pm5$(stat.)$\pm8$ (syst.)\% of the expectation. This proves that the final state 
wave function retains amplitudes for all possible decays, long after the decay occurs.

STAR also observed production of the $\pi^+\pi^-\pi^+\pi^-$ state which can be attributed to the overlap of the two 
separate states $\rho(1450)$ and $\rho(1700)$.

\section*{Acknowledgments} 
This work was in part supported by College of Arts in Science of Creighton University.

\end{document}